\documentclass[12pt]{article}
\usepackage{amsmath}
\usepackage{graphicx}
\usepackage{hyperref}
\usepackage[latin1]{inputenc}
\usepackage{latexsym,amsmath}
\usepackage{hyperref}
\usepackage{amsfonts}
\usepackage{color}
\usepackage{amsmath}
\usepackage{breqn}
\usepackage{dsfont}
\usepackage{enumerate}
\usepackage{natbib}
\usepackage{graphics}

\begin{document}
\title{GRAVITATIONAL WAVE FROM DOMAIN WALLS IN $f(G)$ THEORY }
\author{\small S.P.HATKAR$^1$,S.P.SARAOGI $^2$, S.D.KATORE $^2$ \\
\small $^2$ Department of Mathematics, Sant Gadge Baba Amravati University\\
	\small  AMRAVATI-444602,INDIA\\
	\small E-mail:katore777@gmail.com\\
	\small $^1$ Department of Mathematics, A.E.S. Arts, Commerce and Science\\
	\small COLLEGE, HINGOLI-431513, INDIA\\
	\small E-mail:schnhatkar@gmail.com}

\maketitle
      \textbf{Abstract:} In this paper, we have studied Bianchi type I space-time in the presence of domain walls in the context of $f(G)$   theory of gravitation.  Field equations are solved by using the special form of deceleration parameter. It is also assumed that expansion is proportional to the shear scalar of the model. Some physical parameters are discussed in detail.\\
      
\maketitle
 
 \section{Introduction}
 Modified theories of gravitation (MTG) are proposed to explain accelerated expansion of the universe. It is alternative way to dark energy hypothesis. Late time acceleration, Solar system test, inflation have been sucessfully achieved in $f(R)$ MTG [1-4]. The galatic dynamic of massive test particle in the framework of $f(R)$ theory was constructed by Capozziello et al.[5]. Harko et al.[6] have proposed another extensiontheory,viz. $f(R,T)$ theory, where the gravitational Lagrangian is given by an arbitrary function of the Ricci scalar $R$ and of the trace of the energy tensor $T$. An upper limit on the magnitude of solar system is obtained in $f(R,T)$ MTG. Further,  it is warned that the production of $f(R,T)$ MTG may lead some major differences as compared to that of general theory of relativity. One more developement of general theory of relativity is proposed by introducing Lagrangian is function of $G$, the Gauss-Bonnet curvature invariants. The term $G$ in Lagrangian can avoid ghost contribution and help in regulation of the gravitational action [7]. It is possible to construct cosmologically viable models in $f(G)$ which are consistent with general relativity.
 
 It is well known that universe at early time bahaved like anisotropic.Bianchi type I Kasner form is an important anisotropic solution of general relativistic model. Bianchi type I is simplest generalization of Friedmann-Robertson-Walker model. It is useful to discuss the phenomenon of galaxy formation in early epoch of the universe. There have been numerous studies of Bianchi type I metric in the literature.\\ Katore et al.[8] have explored Bianchi type I for strange quark matter. Pradhan et al.[9] have studied massive string cosmological models in Bianchi type I universe. Yadav et al.[10] have solved Einstein's field equations for Bianchi type I space time. Moreover, the way in which the Bianchi type play role in the investigation of anisotropies makes it the obvious starting points for understanding to cosmological solutions of MTG. In addition to this, topological defects like monopoles, strings and domain walls are formed in the early phase transition of the universe.  Hill et al.[11] have shown that domain walls are important to explain large scale structure of the universe. Decrease of energy density of domain walls is slower than that of matter, therefore, domain walls dominate the universe. If domination of walls survived until the present, it will deteriorate the accomplishment of standard cosmological model [12]. Besides, it will affect the expansion of  the universe, formation of galaxies and excessive anisotropy  in the cosmic microwave background [13]. In this paper we propose to investigate production of gravitational wave from domain walls. Solution of field equations is given in section 2. Section 3, contains a conclusion of our results.
 
 \section{Solution of field equations}

The topological invariant $G$ in four dimension may lead interesting effect and for specific choice of the function $f(G)$ may describe late time acceleration of the universe. Moreover, the $f(G)$ theory have passes solar system test. Past deceleration to recent acceleration transition phase of the universe is also possible in this theory [14]. Let us consider the action of the $f(G)$ gravity as
\begin{equation}
	S_1=\frac{1}{2k}\int[R+f(G)]\sqrt{-g}d^4x + s_\varphi(g^{ij},\varphi),
\end{equation}
where $g$ is the determinant of the metric tensor $g_{ij}$, $k^2= 8\pi G_M, G_M$ is the gravitational constant.  $\varphi$ is the matter field and $s_\varphi$ represents matter action. $f(G)$ is purely metric theory since the matter and metric tensor $g_{ij}$ are minimally coupled. The Gauss-Bonnet invariant $G$ is given by
\begin{equation}
	G=R^2-4R_{ij}R^{ij} + R_{ij\mu\nu}R^{ij\mu\nu},
\end{equation} 
where $R$ and $R_{ij}$ are the Ricci scalar and  Ricci tensor respectively. $R_{ij\mu\nu}$ stands for Riemannian tensors. By  varying the action in equation (1) with respect to the metric $g_{ij}$ ,corresponding field equations are obtained as follows:
\begin{equation}
\begin{split}	
	R_{ij}-\frac{1}{2}R g_{ij}+\delta[R_{i\mu j\nu}+R_{\mu j}g_{\nu i}-R_{\mu\nu}g_{ji}-R_{ij}g_{\nu\mu}+R_{i\nu}g_{j\mu}\\ +\frac{1}{2}R(R_{ij}g_{\mu\nu}-g_{i\nu}g_{j\mu})]\nabla^\mu\nabla^\nu+(Gf_G-f)g_{ij}=kT_{ij},
\end{split}
\end{equation}
here $\nabla_\mu$ denotes the covariant derivative and $f_G$ denotes the derivative with respect to $G$ of the function $f$. 
Large-scale structure of the universe is nearly isotropic at the present time. However, it is anisotropic locally which needs to study by analysing the anisotropic models. When $k=0$ in the Friedmann-Robertson-walker model, the resulting perturbation have to be Bianchi type I. The perturbation die away at late time. Exact power law solution for anisotropic universe in Gauss-Bonnet gravity are obatined by Fayaz et al.[15]. Late time accelerating expansion of the universe is investigated by Li et al.[16]. Nojiri et al.[17] have proposed Gauss-Bonnet dark energy. The Bianchi type I line element is considered  as

\begin{equation}
	ds^2=dt^2-A^2dx^2-B^2\left( dy^2+dz^2\right) ,
\end{equation} 
The Gauss-Bonnet invariant $G$ for the line element (4) is obtained as:
\begin{equation}
	G=-8\left[\frac{\ddot{A}\dot{B^2}}{AB^2}+2\frac{\dot{A}\dot{B}\ddot{B}}{AB^2}\right],
\end{equation}
The motion of domain walls is govened by its surface tension. Thickness of domain walls is comparable with the thermal wavelength $\tau^{-1}$. At $\tau =1Gev$ the thickness of walls are very thin [18].The energy momentum tensor of domain walls is taken as
\begin{equation}
	T_{ij}=\rho (g_{ij}+u_iu_j)+P u_i u_j
\end{equation}
with $u^iu_i=-1$. Recently, Sahoo and Mishra [19] have studied domain walls in Bimetric theory. Katore et al.[20] have investigated domain walls in $f(R,T)$ theory of gravitation. Caglar and Aygun [21] have studied quark matter attached to domain walls in self creation theory.  The field equations (3) for the line element (4) are explicitly written as
\begin{equation}
	\frac{\dot{B^2}}{B^2}+2\frac{\ddot{B}}{B}-16\frac{\dot{B}\ddot{B}}{B^2}\dot{f_G}-8\frac{\dot{B^2}}{B^2}\ddot{f_G}+Gf_G-f= k\rho
\end{equation}     
  
 \begin{equation}
 	\frac{\ddot{B}}{B}+\frac{\ddot{A}}{A}-8\left( \frac{\dot{B}\ddot{A}}{AB}+\frac{\ddot{B}\dot{A}}{BA}\right) \dot{f_G}-8\frac{\dot{B}\dot{A}}{AB}\ddot{f_G}+Gf_G-f= k\rho
 \end{equation}  
  \begin{equation}
  	\frac{\dot{B^2}}{B^2}+2\frac{\dot{A}\dot{B}}{AB}-24\frac{\dot{A}\dot{B^2}}{AB^2}\dot{f_G}+Gf_G-f= -kP
  \end{equation} 
 We have system of three equations (7)-(9) in five unknown $A,B,f,\rho, P$. We need two more conditions to solve the system completely.Firstly we consider physical viable relation between expansion scalar and shear scalar and secondly we assume special form of deceleration parameter which are defined below.
 We define expansion scalar and shear scalar as
 \begin{equation}
 	\theta= \frac{\dot{A}}{A}+2\frac{\dot{B}}{B}
 \end{equation}
\begin{equation}
	\sigma=\frac{1}{\sqrt{3}}\left( \frac{\dot{A}}{A}-\frac{\dot{B}}{B}\right)
\end{equation} 
 Now from the observations of extra galatic sources, it is found that Hubble expansion of the universe may be isotropic when $\frac{\sigma}{\theta}$ is constant [22]. Collins et al.[23]have shown that $\sigma$ is proportional to $\theta$. From equations (10) and (11), we deduce the following relation by assuming that $\sigma$ is proportional to $\theta$. 
 \begin{equation}
 	A=B^n
 \end{equation} 
where $n$ is constant.Shamir and Raza [24] have solved system of $f(R)$ gravity equations by assuming $\frac{\sigma}{\theta}=constant$ for Bianchi type I metric. Chirde et al.[25] have assumed this condition to solve field equations of $f(T)$ gravity for Bianchi type I metric.As discussed above the universe exhibits phase transition from decelerating to accelerating. It is also observed from SNe Ia [26] and CMB anisotropies [27]. Sign of deceleration parameter is used to describe the nature of the universe. It is geometric and positive value indicate deceleration whereas negative value stands for acceleration of the universe. The law of variabtion of Hubble parameter yield a constant value of deceleration parameter which is not consistent with observations. To overcome this, Kumar and Singh [28], Akarsu and Kilinc [29], Katore et al.[30] have proposed different form of deceleration parameter.Keeping this in view, we consider another possibility of deceleration parameter as follows:
\begin{equation}
	q=-\frac{\ddot{S}S}{\dot{S^2}}=-b(t)-1
\end{equation}
where $S^3=V=AB^2$. We choose $b(t)= \frac{t^2}{4}(1-2logt)$, then scale factor $S$ and deceleration parameter read as
\begin{equation}
	S=logt
\end{equation}
\begin{equation}
	q=\frac{t^2}{4}(2logt-1)-1
\end{equation}
\begin{figure}
	\includegraphics[width=16cm] {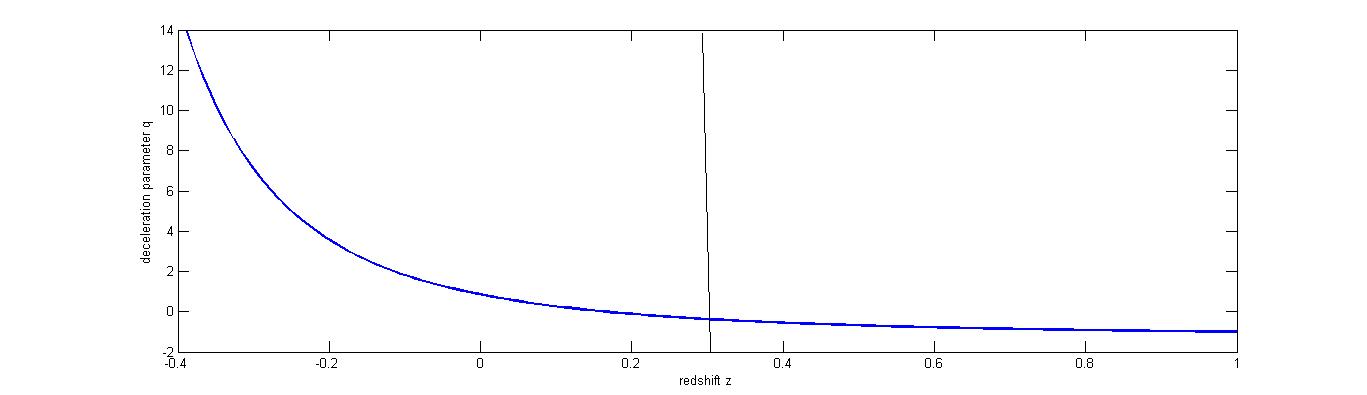}
	\caption{Plot of deceleration parameter versus redshift z}
\end{figure}  
Figure (1) reveals that the sign of deceleration parameter is negative in the past $(z>0)$ and at present $(z=0)$ and positive in the future $(z<0)$. This shows that the universe was accelerating at present and decelerating in the future.
 Using equations (12) and (14) we obtain the follwoing expression of metric potentials: 
 \begin{equation}
 	A= \left(logt \right)^{\frac{3n}{n+2}} 
 \end{equation} 
 \begin{equation}
 	B= \left(logt \right)^{\frac{3}{n+2}} 
 \end{equation}  
  The expansion scalar $(\theta)$ and shear scalar $(\sigma)$ are found to be
  \begin{equation}
  	\theta= \frac{3}{tlogt}
  \end{equation}
  \begin{equation}
  	\sigma= \frac{sqrt{3}(n-1)}{(n+2)tlogt}
  \end{equation}
Figure (2) shows that the rate of expansion of the universe is decreasing function. It was large in the past. The $\sigma\rightarrow 0$ as $z\rightarrow -1$ i.e. the initial anisotropic universe tends to isotropic one. It is remarkable to note that it is as per expectation.
\begin{figure}
	\includegraphics[width=16cm] {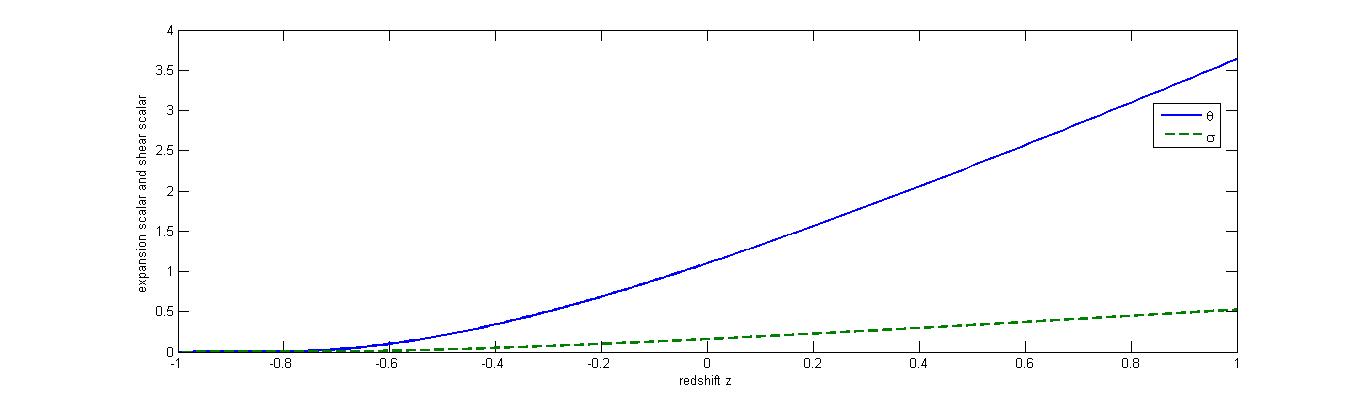}
	\caption{Plot of expansion scalar and shear scalar versus redshift z}
\end{figure}

\begin{figure}
	\includegraphics[width=16cm] {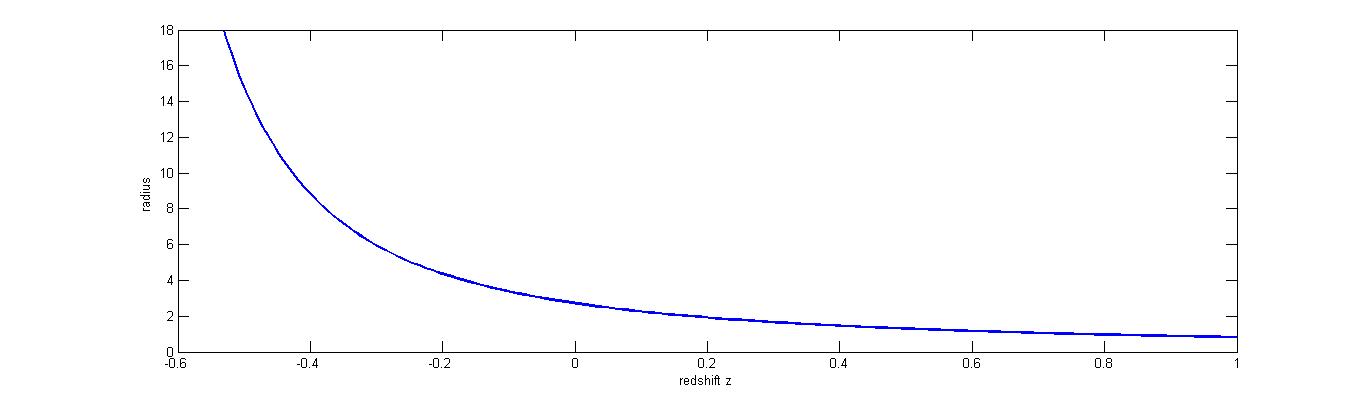}
	\caption{Plot of Radius of domain walls versus redshift z}
\end{figure}
  The Gauss Bonnet Invariant ($G$) is obtained as
  \begin{equation}
  	G=-\frac{648n}{(n+2)^3t^4(logt)^3}
  \end{equation}
 Using equations (7),(8),(16),(17), (20) the Gauss Bonnet function in terms of time is obtained as
 \begin{equation}
 	f=-\frac{l_7}{t^2(logt)^3}+\frac{l_8}{t^4(logt)^2}+\frac{l_9}{t^4(logt)^3}\sum_{m=1}^{\infty}\frac{(2t)^m}{m\times m!}
 \end{equation} 
  where $l_1=\frac{n^2+n-2}{72(n-1)}+\frac{7(n+2)}{864},l_2=\frac{n^2+n-2}{108(n-1)}+\frac{(n+2)}{324},l_3=\frac{n^2+n-2}{108(n-1)}+\frac{2}{27},l_4=\frac{n^2+n-2}{36(n-1)}+\frac{(n+2)}{108},l_5=\frac{(n+2)}{72}, l_6=\frac{n^2+n-2}{108(n-1)}+\frac{(n+2)}{324}, l_7=\frac{648nl_1}{(n+2)^3}, l_8=\frac{648nl_2}{(n+2)^3},l_9=\frac{648nl_3}{(n+2)^3}$ \\
  The energy density is obtained as
  
  \begin{equation}
  	\begin{split}
  	k\rho=\frac{d_1}{t}-\frac{d_2}{tlogt}-\frac{d_3}{t(logt)^2}+\frac{d_4}{(tlogt)^2}+\frac{d_5}{t^2logt}+\frac{d_6}{t^2(logt)^3}\\ +\frac{d_7}{t^2(logt)^4}+\frac{d_8}{t^4(logt)^2}+\frac{d_9}{t^4(logt)^3}\sum_{m=1}^{\infty}\frac{(2t)^m}{m\times m!}
  \end{split}
    \end{equation}
  
 where $d_1=\frac{24l_5}{n+2}, d_2= \frac{24(l_4-l_5)}{n+2}, d_3= \frac{24l_6}{n+2}, d_4= \frac{15-6n}{(n+2)^2}+\frac{144l_4}{(n+2)^2}+\frac{144(1-n)l_5}{(n+2)^3},, d_5= -\frac{6}{n+2}-\frac{144l_5}{(n+2)^2}, d_6= -\frac{144(1-n)l_4}{(n+2)^3}-\frac{144l_6}{(n+2)^2}+l_7 -\frac{648nl_1}{(n+2)^3}, d_7=\frac{144(1-n)l_6}{(n+2)^3}, d_8= \frac{648nl_2}{(n+2)^3}-l_8, d_9= \frac{648nl_3}{(n+2)^3}-l_9$ \\
  The pressure is obtained as
  \begin{equation}
  		-kP=\frac{h_1}{(tlogt)^2}+\frac{h_2}{t^2(logt)^3} +\frac{h_3}{t^2(logt)^4}+\frac{h_4}{t^4(logt)^2} \\+\frac{h_5}{t^4(logt)^3}\sum_{m=1}^{\infty}\frac{(2t)^m}{m\times m!}
   \end{equation}
  where $h_1=\frac{648l_5}{(n+2)^3}+\frac{9(1+2n)}{(n+2)^2}, h_2= l_7- \frac{648nl_4}{(n+2)^3}-\frac{648nl_1}{(n+2)^3}, h_3= \frac{648nl_6}{(n+2)^3}, h_4=\frac{648nl_2}{(n+2)^3}- l_8, h_5= \frac{648nl_3}{(n+2)^3}-l_9$ \\
 When the domain walls are formed in the early universe, there is one domain wall per Hubble horizon [12,31]. They follows scaling law as [31] 
 \begin{equation}
 	\rho \sim \sigma_d H
 	\end{equation}
 where $\sigma_d$ is the tension density of domain walls and $H=\frac{1}{3}\theta$ is Hubble parameter.
  The tension density $(\sigma_d)$ of the domain wall is obtained as
  	\begin{equation}
  	\begin{split}	
   	k\sigma_d=d_1logt- d_2- \frac{d_3}{(logt)}+\frac{d_4}{(tlogt)}+\frac{d_5}{t}+\frac{d_6}{t(logt)^2}\\
   	 +\frac{d_7}{t(logt)^3}+\frac{d_8}{t^3(logt)}+\frac{d_9}{t^3(logt)^2}\sum_{m=1}^{\infty}\frac{(2t)^m}{m\times m!}
   	 \end{split}
 \end{equation}
Now we define $H_d$ as[32] 
\begin{equation}
	H_d\equiv \frac{\sigma_d}{M_p^2}
\end{equation}
where $M_p$ denotes reduced Plank scale. At $H\sim H_d$, the domain walls starts to dominate and the universe will be extremely inhomogeneous afterwards.The radius of wall $R_d$ and the distance of two beighbouring walls $L_d$ are defined as [33]
\begin{equation}
	R_d\sim L_d\sim H^{-1}
\end{equation}
The radius of domain walls was small in the past $z>0$ than the present $z=0$ and it is increasing with decreasing redshift. In the future (Z<0), it will be large. The distance between two neighbouring domain wall is increasing. It was small in the past and tends to large in the future (see figure 3).

From figure (4), it is clear that energy density of walls was large in the past and tends to zero as $z\rightarrow -1$. The tenstion density of the wall was small in the past and diverges with decreasing redshift i.e. it means that walls will be not present in the future, it is according to the expectation.

It has been pointed out that when radius of wall was small the tension density was also small whereas energy density was large in the past. As the radius increases, the tension density also increasing and energy density decreses with decreasing redshift.
We have adopted the formulation of Nakayama et al.[32],let us roughly estimate the energy density of gravitational wave (gw).The power of gravitational wave (gw) is obtained from the  following quadrupole radiation formula:
\begin{equation}
\dot{E_{gw}}\sim \frac{1}{M_p^2}\left( \frac{d^3J}{dt^3}\right)^2
\end{equation} 	
 where $J$ stands for quadrupole moment of the objects. In the scaling regime quadrupole moments is obtained to be $J\cong\frac{\sigma_d}{H^4}$. Then the gravitational wave emitted during one Hubble time is given by
 \begin{equation}
 	\rho_{gw}\sim\frac{\dot{E_{gw}}H^{-1}}{H^{-3}}\sim\frac{\sigma_d}{M_p^2}
 \end{equation}
 From figure (5), it is observed that energy density of gravitational wave was large in the past as compare to the present. It will be large in the future. As the tension of domain walls is increasing from certain value of redshift, the domain walls must be vanished. There is bound on the tension of domain walls.$\sigma_d>O (MeV^3)$,[34]. Due to the existence of the energy bias, domain walls may anihilate before vanish and produce gravitational waves [34].
  \begin{figure}
  	\includegraphics[width=16cm] {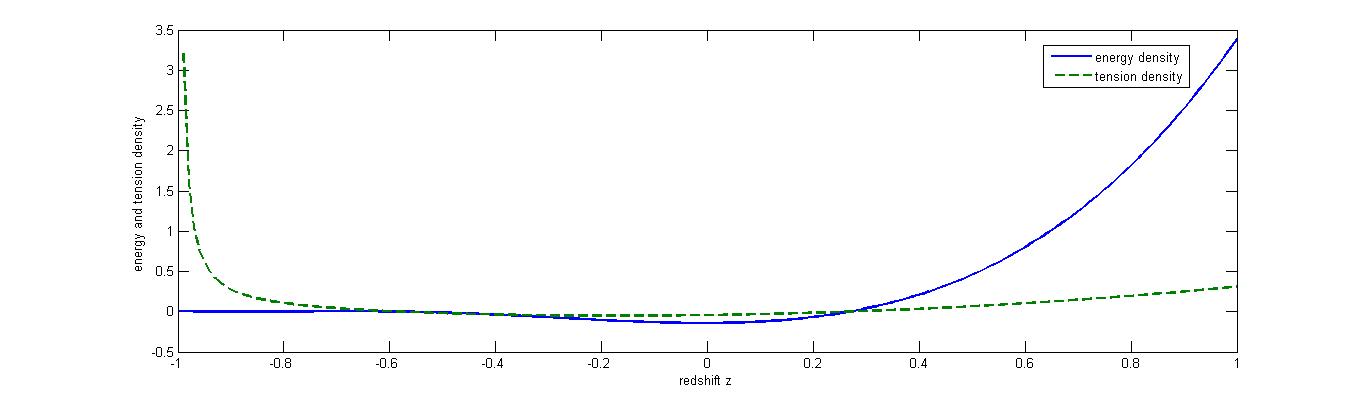}
  	\caption{Plot of energy density and tension density versus redshift z}
  \end{figure}  

   \begin{figure}
 	\includegraphics[width=16cm] {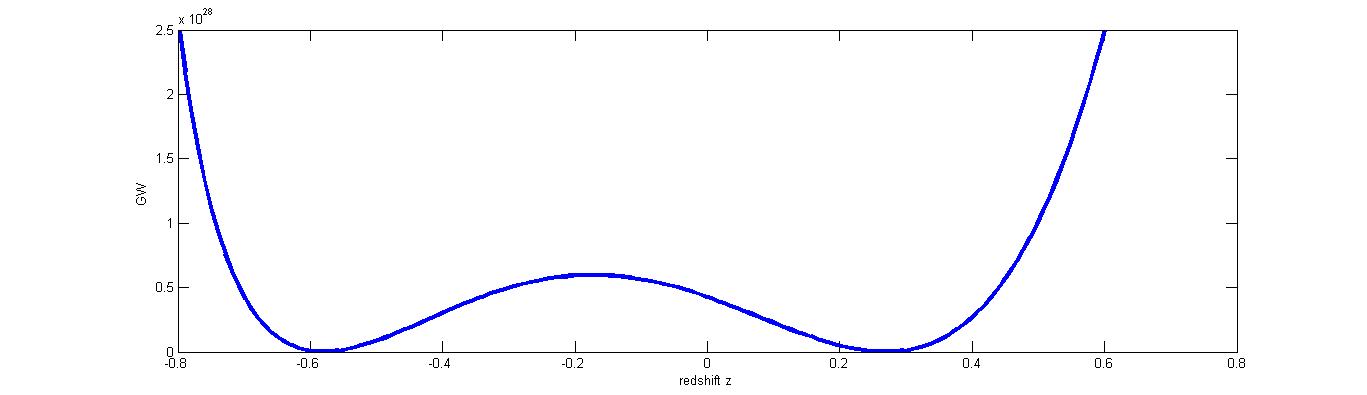}
 	\caption{Plot of energy density of gravitational wave versus redshift z}
 \end{figure} 

 \section{Conclusion}
 In this papaer we have presented Bianchi type I space time in the context of $f(G)$ theory of gravitation for domain walls. Production of gravitational wave from domain walls is estimated. It is found that the universe is accelerating and expanding at the present. The anisotropy of the universe is vanished at the present i.e. the universe is isotropic. The radius of domain walls was small in the past and it is increasing with decreasing redshift. The energy density of the domain walls is decreasing function of redshift. Tension density of domain wall is increasing with decreasing redshift. The domain walls are not present The distance between two neighbouring domain walls is also increasing with decreasing redshift. The energy density of gravitation wave is large in the past and in the future comapare to the present. It may possible that they anihiated, collaps and produced gravitational wave.
 \section{References} 
1. S.M. Carroll, V.Duvvuri, M. Trodden, M.S. Turner, Phys.Rev.D. 70,043528,2004.\\
2.W. Hu, I. Sawick, Phys.Rev.D.76,064004,2007.\\
3.S. Nojiri, S.D. Odintsov, Phys.Rev.D.68,123512,2003.\\
4.S. Nojiri, S.D. Odintsov, Phys. Lett. B.657,238,2007.\\
5.S. Capozziello, V.F. Cardone, A. Troisi, Mon. Not.R. Astron.Soc. 375,1423,2007.\\
6.T. Harko, F.S.N.Lobo, S. Nojiri, S.D. Odintsov, arXiv:1104.2669v2,2011.\\
7.M.De Laurentis, A. J. Loper-Revelles, arXiv:1311.0206v1,2013.\\
8.S.D.Katore, S. P. Hatkar, R. J.Baxi, Found. Phys.46,409-427,2016.\\
9. A. Pradhan, S.S. Kumhar, K. Jotania, Palestine. J.Math. 1(2),117-132,2012.\\
10 P.Yadav, S.A. Faruqui, A. Pradhan, ARPN J. Sci. Tech.3,7,2013.\\
11.C.T. Hill, D.N. Schramm, J.N. Fry,Comments Nucl. Part. Phys.19,25,1989.\\
12.W.H.Press, B.S.Ryden, D.N. Spergel, Astrophys.J.347,590,1989.\\
13.G.B. Gelmini, M. Gleiser, E.W.Kolb, Phys. Rev.D.39,1558,1989.\\
14.S. Nojiri, S.D.Odintsov,Phys. Lett. B.631,1-6,2005.\\
15.V. Fayaz, H. Hossienkhani, A. Aghamohammadi, Astrophys.Space Sci.357,136,2015.\\
16.B.Li, J.D. Barrow, D.F.Mota, arXiv:0705.3795v3,2007.\\
17.S. Nojiri, S.D.Odintsov,M.Sasaki, arXiv:hep-th/0504052v2,2005.\\
18.T.W.B.kibble, J.Phys. A:Math, Gen. 9,8,1976.\\
19.P.K.Sahoo, B. Mishra, The Afric. Rev.Phys.8,0053,2013.\\
20.S.D.Katore, S.P. Hatkar, R. J. Baxi, Chin. J. Phys.\\
21.H. Caglar, S. Aygun, Chin. Phys.C.40,4,045103,2016.\\
22.R. Kantowski, R.K. Sachs,J.Math.Phys.7,433,1966.\\
23.C.B.Collins, E.N.Giass, D.A. Wilkinson, Gen.Relativ.Gravit.13,805,1980.\\
24.M.F.Shamir, Z.Raza, Can.J.Phys.93,1-6,2015.\\
25.V.R.Chirde, S.P.Hatkar, S.D.Katore, Int.J. Mod.Phys.D.29,8,2050054,2020.\\
26. A.G. Riess et al.Astron.J.116,1009,1998.\\
27.M.S. Berman, Nuovo.Cimento B 74,182,1983.\\
28.S.Kumar, C.P.Singh, Astrophys.Space Sci.312,57-62,2007.\\
29.O.Akarsu, C.B.Kilinx, Gen. Relativ.Gravit.42,763,2010.\\
30.S.D.Katore, S.P.Hatkar, P.S.Dudhe, Astrophys.64,1,103-116,2021.\\
31. T. Garagounis, M.Hindmarsh, Phys. Rev.D.68,103506,2003.\\
32.K.Nakayama, F.Takahashi, N. Yokozaki, arXiv:1612.08327v1,2016.\\
33.T. Hiramatsu, M.Kawasaki, K.Saikawa, arXiv:1002.1555v2,2010.\\
34.K. Saikawa, Universe, 3,40,2017.\\

\end{document}